%% file: VanderSwaelmenEtAl2015.tex
\documentclass[structabstract]{aa} 

\usepackage{savesym}

\usepackage{xspace}

\usepackage{epstopdf}
\usepackage{graphicx}

\usepackage[table]{xcolor}
\usepackage{bigstrut}
\usepackage{booktabs}
\usepackage{pdflscape}
\usepackage{longtable}
\usepackage{multirow}

\usepackage{array}
\newcolumntype{+}{>{\global\let\currentrowstyle\relax}}
\newcolumntype{^}{>{\currentrowstyle}}

\usepackage[a4paper,colorlinks=true,pdffitwindow=true,pdfpagelayout=SinglePage,pdfpagemode=UseOutlines,bookmarks=true,bookmarksopen=true,bookmarksnumbered=true]{hyperref}

\usepackage{amsmath}

\savesymbol{iint} 
\usepackage{txfonts} 
\restoresymbol{TXF}{iint} 

\DeclareSymbolFont{TXFontLetterFix}{OML}{cmm}{m}{it}
\DeclareMathSymbol{\Substitutev}{\mathalpha}{TXFontLetterFix}{118}

\usepackage{natbib} 
\bibpunct{(}{)}{;}{a}{}{,} 

\usepackage[printonlyused,withpage]{acronym} 

\usepackage[normalem]{ulem}
\usepackage{pifont}

\usepackage{mhchem}

\savesymbol{arcmin}
\savesymbol{arcsec}
\savesymbol{fs}
\usepackage{siunitx}
\restoresymbol{SIUX}{arcmin}
\restoresymbol{SIUX}{arcsec}
\restoresymbol{SIUX}{fs}
\DeclareSIUnit\Year{yr}
\DeclareSIUnit\dex{dex}
\DeclareSIUnit\parsec{pc}
\DeclareSIUnit\magnitude{mag}
\DeclareSIUnit\lines{lines}
\DeclareSIUnit\ElectronUnit{\text{\ensuremath{\mathrm{e^{-}}}}}
\DeclareSIUnit\PixelUnit{pix}
\DeclareSIUnit\PhotonUnit{photon}

\newcommand{\abratio}[2]{[\mathrm{#1}/\mathrm{#2}]\xspace}
\newcommand{\ElementDegree}[2]{#1\,#2\xspace}

\newcommand{\MODIFter}[1]{#1\xspace}

\hypersetup
    { 
      pdftitle=Heavy elements Ba\, La\, Ce\, Nd\, and Eu in 56 Galactic bulge red giants,   
      pdfauthor=M. Van der Swaelmen; B. Barbuy; V. Hill; M. Zoccali; D. Minniti; S. Ortolani; A. G{\'o}mez,   
      pdfsubject=Astrophysics,   
      pdfcreator=,
      pdfproducer=,
      pdfkeywords=stars: abundances\, atmospheres - Galaxy: bulge,   
    }

\begin{document}

\acrodef{hfs}{hyperfine structure}
\acrodef{RGB}[RGB]{red giant branch}
\acrodef{AGB}[AGB]{asymptotic giant branch}
\acrodef{CMD}[CMD]{colour-magnitude diagram}
\acrodef{SNII}[SNII]{type II supernova}
\acrodef{SNIa}[SNIa]{type Ia supernova}
\acrodef{SNeII}[SNII]{type II supernovae}
\acrodef{SNeIa}[SNIa]{type Ia supernovae}
\acrodef{ISM}[ISM]{intestellar medium}
\acrodef{IMF}[IMF]{initial mass function}

\title{Heavy elements Ba, La, Ce, Nd, and Eu in 56 Galactic bulge red giants\thanks{Observations collected at the European Southern Observatory, Paranal, Chile (ESO programmes 71.B-0617A, 73.B0074A, and GTO 71.B-0196)}}

\author{M. Van der Swaelmen\inst{\ref{institute1}}
  \and B. Barbuy\inst{\ref{institute1}}
  \and V. Hill\inst{\ref{institute2}}
  \and M. Zoccali\inst{\ref{institute3},\ref{institute4}}
  \and D. Minniti\inst{\ref{institute4},\ref{institute5},\ref{institute6}}
  \and S. Ortolani\inst{\ref{institute8}}
  \and A. G\'omez\inst{\ref{institute9}}
}

\offprints{M. Van der Swaelmen}

\institute{Universidade de S\~ao Paulo, IAG, Rua do Mat\~ao 1226, Cidade Universit\'aria, S\~ao Paulo 05508-900, Brazil\\
  \email{mathieu.vanderswaelmen@iag.usp.br}\label{institute1}\\
  \email{b.barbuy@iag.usp.br}
  \and Laboratoire Lagrange (UMR7293), Universit\'e de Sophia-Antipolis, CNRS, Observatoire de la C\^ote d'Azur, CS 34229, 06304 Nice Cedex 4, France\\
  \email{vanessa.hill@oca.eu}\label{institute2}
  \and Instituto de Astrofisica, Facultad de Fisica, Pontificia Universidad Catolica de Chile, Casilla 306, Santiago 22, Chile\\
  \email{mzoccali@astro.puc.cl}\label{institute3}
  \and Millennium Institute of Astrophysics, Casilla 306, Santiago 22, Chile\label{institute4}\\
  \email{dante@astrofisica.cl}
  \and Departamento de Ciencias Fisicas, Universidad Andres Bello, Republica 220, Santiago, Chile\label{institute5}
  \and Vatican Observatory, V00120 Vatican City State, Italy\label{institute6}
  \and Universit\`a di Padova, Dipartimento di Astronomia, Vicolo dell'Osservatorio 2, I-35122 Padova, Italy\\
  \email{sergio.ortolani@unipd.it}\label{institute8}
  \and Observatoire de Paris-Meudon, 92195 Meudon Cedex, France\\
  \email{anita.gomez@obspm.fr}\label{institute9}
}

\date{submitted 21/01/2015; accepted 28/09/2015}

\abstract{}
         {The aim of this work is the study of abundances of the heavy elements Ba, La, Ce, Nd, and Eu in $56$ bulge giants (red giant branch and red clump) with metallicities ranging from \SI{-1.3}{\dex} to \SI{+0.5}{\dex}.}
         {We obtained high-resolution spectra of our giant stars using the FLAMES-UVES spectrograph on the Very Large Telescope. We inspected four bulge fields along the minor axis.}
         {We measure the chemical evolution of heavy elements, as a function of metallicity, in the Galactic bulge.}
         {The $\abratio{Ba, La, Ce, Nd}{Fe}$ vs. $\abratio{Fe}{H}$ ratios decrease with increasing metallicity, in which aspect they differ from disc stars. In our metal-poor bulge stars, La and Ba are enhanced relative to their thick disc counterpart, while in our metal-rich bulge stars La and Ba are underabundant relative to their disc counterpart. Therefore, this \MODIFter{contrast between bulge and discs trends} indicates that bulge and (solar neighbourhood) thick disc stars could behave differently. An increase in $\abratio{La, Nd}{Eu}$ with increasing metallicity, for metal-rich stars with $\abratio{Fe}{H} > \SI{0}{\dex}$, may indicate that the \emph{s}-process from AGB stars starts to operate at a metallicity around solar. Finally, $\abratio{Eu}{Fe}$ follows the $\abratio{\alpha}{Fe}$ behaviour, as expected, since these elements are produced by SNe type II.}


\authorrunning{M. Van der Swaelmen et al.}
\titlerunning{Heavy elements in Galactic bulge red giants}

\maketitle

\section{Introduction} 
\label{Sec:Introduction}

Heavy elements (with $Z > 30$) are produced by neutron captures through \emph{s}- and \emph{r}-processes in a variety of sites. In strict terms, the \emph{s}-process indicates a situation where no neutron captures take place on $\beta$-unstable nuclei, while the \emph{r}-process indicates a situation where neutron captures occur before a radioactive $\beta$-decay. Different channels have been identified for s-element production \citep{1989RPPh...52..945K,2011RvMP...83..157K,1999ARA&A..37..239B,2012A&A...538L...2F,2014ApJ...787...10B}: the weak component, contributing elements of $A \leq 90$, takes place during He core and convective shell C burning of massive stars; and the main and strong components ($A \sim 90-208$) are produced in helium rich intershells of thermally pulsing \ac{AGB} stars \citep{2011MNRAS.418..284B,2014ApJ...787...10B}. According to, for example, \cite{1999ApJ...525..886A,2004ApJ...617.1091S}, the elements Ba, La, Ce, and Nd in the solar neighbourhood, are dominantly produced by the \emph{s}-process. However, \cite{1981A&A....97..391T} suggested that, in very old stars, the abundances of these elements are the result of their \emph{r}-process fraction produced at early times. This can apply to the Galactic bulge, which shows its bulk of stars as being very old \citep{2003A&A...399..931Z,2008ApJ...684.1110C}. \MODIFter{Of the 58 dwarf stars observed thanks to microlensing effects, \cite{2013A&A...549A.147B} estimated that \SI{22}{\percent} are younger than \SI{5}{\giga\Year} and \SI{38}{\percent} are younger than \SI{7}{\giga\Year} (if He is standard)}, and these stars are among the more metal-rich stars. If a higher value of the He content is chosen, then age determination leads to a smaller fraction of stars younger than $6-\SI{8}{\giga\Year}$. The old bulge stars could have their heavy element abundances produced by the \emph{r}-process, as seems to be the case for metal-poor halo stars (\emph{e.g.} \citealp{2012A&A...548A..42S}).

The \emph{r}-process is less well understood than the \emph{s}-process, and there is still no consensus as to where the \emph{r}-process takes place, as discussed in, for example, \mbox{\cite{2006NuPhA.777..676W}}; \mbox{\cite{2007ApJ...662...39K}}; \cite{2011LNP...812..153T}; \cite{2013ENews..44...23L}; \cite{2014MNRAS.439..744R}, among others. Recently, the hypothesis of \emph{r}-element nucleosynthesis taking place in neutron star merger events is favoured by \cite{2014MNRAS.439..744R} and \cite{2014ApJ...789L..39W}. These authors suggest that the strong \emph{r}-process produces elements with $A > 130$, and the weak \emph{r}-process yields elements with $50 < A < 130$. Finally, massive spinstars can produce \emph{s}-elements very early in the Galaxy, as proposed by \cite{2011Natur.472..454C,2013AN....334..595C}.

The centre of our Galaxy hosts a dense stellar structure, and there is a strong debate concerning its formation process. The Galactic bulge shares chemical and kinematical properties compatible with a classic spheroid and a boxy bulge \citep{1996ApJ...459..175M}. Recently, \cite{2010A&A...519A..77B}, \cite{2014A&A...563A..15B}, \cite{2011A&A...534A..80H}, \cite{2013A&A...549A.147B} and \cite{2014A&A...569A.103R} have been able to identify two different stellar populations using chemical and kinematical properties of stellar samples. They found a stellar population that is old, metal-poor, alpha-element-rich, and has isotropic kinematics, and a younger population that is metal-rich with solar $\alpha$-ratios and bar-like kinematics. Chemically and kinematically speaking, the metal-poor component has similarities with the thick disc, while the metal-rich component is more similar to an inner thin disc. \cite{2013MNRAS.430..836N} carried out a large survey of bulge stars, and were able to separate up to five stellar populations in terms of kinematics and metallicity distributions. They also suspect that the different bulge components have an origin in the thin and thick discs of the Galaxy.

A number of scenarios have been described to explain how to form a spheroidal bulge. For instance, \cite{1962ApJ...136..748E}, \cite{1999ApJ...514...77N}, or \cite{2008ApJ...688...67E} form a spheroidal bulge by gravitational collapse or hierarchical merging of subclumps formed in the disc. On the other hand, \cite{2005A&A...436..127J} or \cite{2010AJ....140.1719S} are able to form a boxy bulge via heating processes after the formation of a dynamical bar due to disc instabilities. \MODIFter{However, there is no consistent explanation of all the observational features of the bulge.} This is expected, given that our Galaxy is an intermediate Sbc type spiral galaxy, and the bulges are considered to be true bulges in Sa and Sb type galaxies, and boxy bulges forming from the bar in Sc type galaxies \citep{2010AJ....140.1719S}.

Chemical abundances of $\alpha$-elements have been used to obtain insights into the chemical history of the Galactic bulge (\emph{e.g.} \citealp{1994ApJS...91..749M}, \citealp{2007ApJ...661.1152F}, \citealp{2006A&A...457L...1Z}, \citealp{2007A&A...465..799L}, \citealp{2011A&A...530A..54G}, \citealp{2011A&A...534A..80H}). In the present work, we try to explore the possibility of using the abundances of heavy elements to characterise the bulge stellar population chemically, and \MODIFter{to provide hints about nucleosynthesis processes and the kind of supernovae (SN) that enriched the Galactic bulge}. We derive heavy element abundances for a sample of \num{56} bulge field stars observed at high spectral resolution with the FLAMES-UVES spectrograph at the Very Large Telescope, as described in \cite{2007A&A...465..799L}, \cite{2006A&A...457L...1Z,2008A&A...486..177Z}.

We compare our results with previous samples in the literature, for which heavy element abundances were derived in Galactic bulge stars. \cite{2012ApJ...749..175J} derived abundances of the heavy elements La, Nd, and Eu in red giants in Plaut's field, among which they found r-II stars \citep{2013ApJ...775L..27J}. \cite{2013A&A...549A.147B} presented element abundances of $58$ microlensed bulge dwarfs and subgiants of the Galactic bulge. Their study includes the abundances of the heavy element Ba.

In Sect.~2 we report our observations. In Sect.~3 we provide the atomic constants for the lines under study. In Sect.~4 we list the basic stellar parameters and describe the abundance derivation of heavy element. The results are derived in Sect.~5 and discussed in Sect.~6. Conclusions are drawn in \MODIFter{Sect.~7}.

\section{Observations}

The spectra we used for this study on heavy elements have already been used for light $\alpha$-elements in \cite{2006A&A...457L...1Z,2007A&A...465..799L} and the iron-peak element Mn and Zn \citep{2013A&A...559A...5B, 2015A&A...580A..40B}. These papers provide further details on the target selection, data reduction, and stellar parameters determination. We summarize these papers below.

\cite{2006A&A...457L...1Z} obtained the spectra of $56$ bulge giant stars with the multifibre spectrograph FLAMES/UVES ($R \sim 45000$) at the UT2 Kuyen VLT/ESO telescope. The spectra, obtained with the UVES spectrograph, span a wavelength range from $\sim 4800$ to $\sim \SI{6800}{\angstrom}$, with a gap (inherent to the instrument design) between $5775$ and $\SI{5825}{\angstrom}$. The targets are located in four different fields of the Galactic bulge: the Baade's Window ($l = \SI{1.14}{\degree}$, $b = \SI{-4.2}{\degree}$), a field at \SI{-6}{\degree} ($l = \SI{0.2}{\degree}$ , $b = \SI{-6}{\degree}$), the Blanco field ($l = \SI{0}{\degree}$, $b = \SI{-12}{\degree}$), and the field of NGC$6553$ ($l = \SI{5.2}{\degree}$, $b = \SI{-3}{\degree}$). Thirteen of our $56$ stars are red clump (RC) stars and located in the Baade's Window, the others are red giant branch (RGB) stars slightly brighter than the RC. \cite{2007A&A...465..799L} performed the data reduction with the ESO FLAMES/UVES pipeline (debiasing, flat-fielding, spectrum extraction, wavelength calibration, order merging), and they performed the sky subtraction and the co-addition of multiple observations with IRAF tasks.

The stellar parameters (temperature $T_{\text{eff}}$, surface gravity $\log g$, overall metallicity $[\text{M}/\text{Fe}]$, and microturbulence velocity $v_{\text{turb}}$) were derived by \cite{2007A&A...465..799L} for the RGB stars and by \cite{2011A&A...534A..80H} for the RC stars. The stellar parameters were derived by applying the standard spectroscopic criteria to their set of iron lines.

\section{Abundances}
\label{Sec:Abundances}

\subsection{Methods}
\label{Sec:Methods}

\begin{figure}
  \begin{centering}
    \includegraphics[width=\columnwidth]{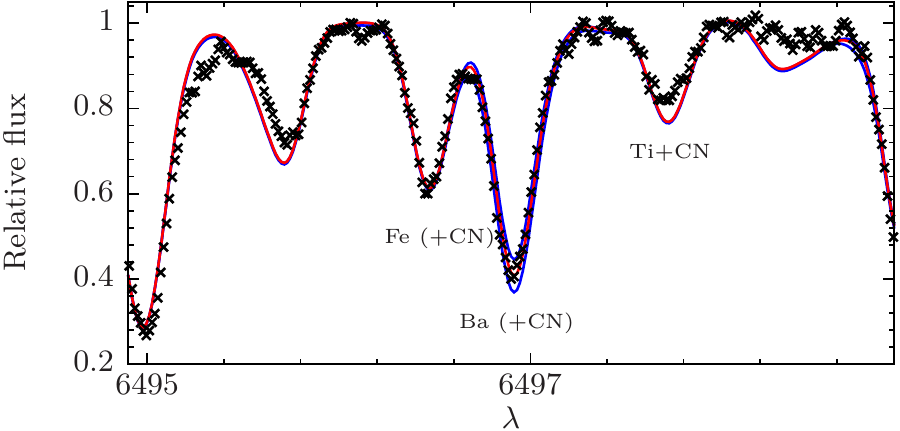}

    \includegraphics[width=\columnwidth]{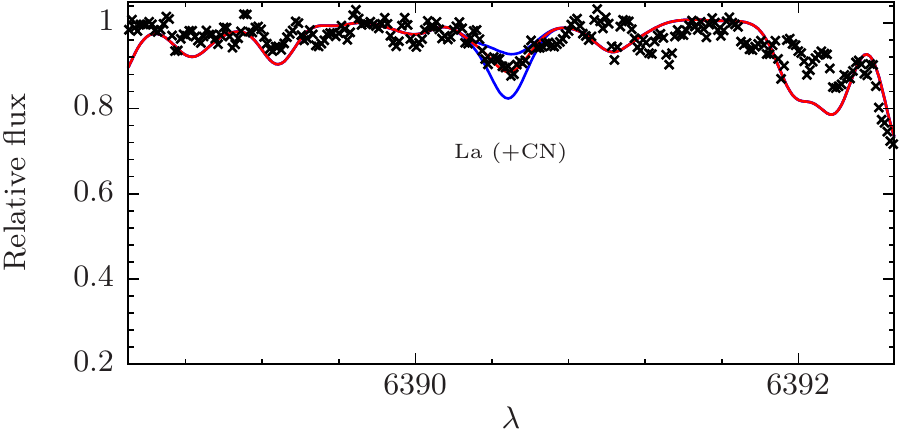}

    \includegraphics[width=\columnwidth]{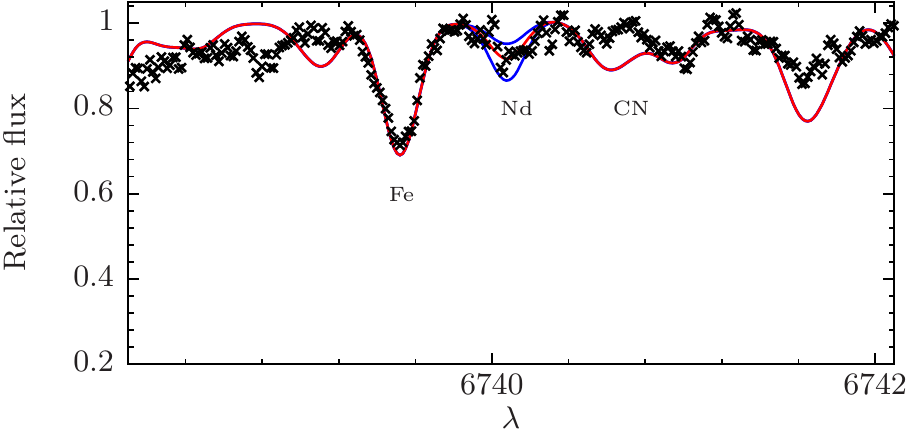}

    \includegraphics[width=\columnwidth]{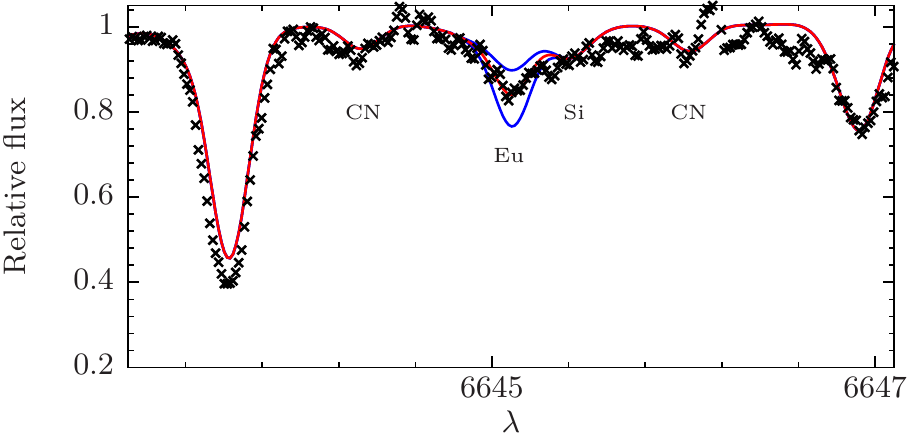}

    \caption{\label{Fig:BWRC04_fits} From top to bottom: fit of \ion{Ba}{II} line at \SI{6496}{\angstrom}, \ion{La}{II} line at \SI{6390}{\angstrom}, \ion{Nd}{II} line at \SI{6740}{\angstrom}, and \ion{Eu}{II} line at \SI{6645}{\angstrom} for BWc-4. Black crosses: observed spectrum; red solid line: best fit; blue solid lines: spectra computed for the best-fit value of $\abratio{X}{Fe} \pm \SI{0.3}{\dex}$.}
  \end{centering}
\end{figure}

We obtained element abundances through line-by-line spectrum synthesis calculations. We used the technique of line fitting described in \cite{2013A&A...560A..44V} to derive the elemental abundances, which is based on a weighted $\chi^{2}$ minimisation with weights inversely proportional to the computed blends from other species at each wavelength. We computed a grid of synthetic spectra using \emph{turbospectrum} (code of line synthesis described in \citealp{1998A&A...330.1109A} and improved along the years by B. Plez) together with the grid of OSMARCS spherical model atmospheres\footnote{models available at \url{http://marcs.astro.uu.se/}} \citep{2008A&A...486..951G}. We computed the synthetic spectra in spherical geometry, with LTE spherical radiative transfer. Our $\chi^{2}$ algorithm then found the best match between the synthetic and observed spectra. Figure~\ref{Fig:BWRC04_fits} shows the fits to the observed spectrum of the RC star Bwc-4 for the lines of Ba, La, Nd, and Eu. Table~\ref{Tab:Abundances_Bulge_stars} gives individual abundances for our bulge stars.

\begin{table*}
  \begin{center}
    \caption{\label{Tab:Abundances_Bulge_stars} Abundances line-by-line derived in the present work.}
    \input{./Table.type=abundances_bulge_stars.tex}
  \end{center}
\end{table*}

\subsection{Line lists}

We compiled the atomic line lists from the line database VALD\footnote{\url{http://www.astro.uu.se/~vald/php/vald.php}} \citep{1999A&AS..138..119K, 2000BaltA...9..590K}. We took into account the \ac{hfs} for \ElementDegree{Ba}{II} (\citealp{1978SoPh...56..237R}: $\SI{6496.912}{\angstrom}$), \ElementDegree{La}{II} (\citealp{2001ApJ...556..452L}: $\SI{6262.287}{\angstrom}$, $\SI{6320.430}{\angstrom}$, $\SI{6390.477}{\angstrom}$), and \ElementDegree{Eu}{II} (\citealp{2001ApJ...563.1075L}: $\SI{6645.064}{\angstrom}$). Table~\ref{Table:loggf} lists the lines we used with their $\log gf$.

\begin{table}
  \begin{center}
    \caption{\label{Table:loggf} Central wavelengths, excitation potential, and adopted oscillator strengths $\log gf$ values, taken from 1 \cite{2013A&A...560A..44V}, otherwise VALD values.}
    \input{./Table.type=lines_data.tex}
  \end{center}
\end{table}

Given the cool effective temperature ($T_{\mathrm{eff}} \sim \SI{4500}{\kelvin}$) of our targets, molecules are expected to form in the atmosphere of our bulge stars and are responsible for molecular bands in the stellar spectra. Therefore, we included the molecular line lists of ${}^{12}\mathrm{C}{}^{14}\mathrm{N}$, ${}^{13}\mathrm{C}{}^{14}\mathrm{N}$ (Plez, private communication) and TiO \citep{1998A&A...337..495P} in the spectrum synthesis.

\subsection{Initial chemical composition}
\label{Sec:Intial_chemical_composition}

We have to set CNO abundances before any abundance derivation since the CNO mix  drives the equilibrium between atomic species, C, N and O, and molecular species, CN and CO, and therefore the CNO mixture controls the strength of CN features. \cite{2007A&A...465..799L} measured individual $\abratio{C}{Fe}$ and $\abratio{N}{Fe}$ ratios for our bulge stars. Since they found no trend with metallicity and since their individual measurements suffer rather large uncertainties, we decided to use mean C and N abundances and we fixed the following values for any star of our sample: $\abratio{C}{Fe} = \SI{-0.05}{\dex}$ and $\abratio{N}{Fe} = \SI{0.35}{\dex}$. \cite{2007A&A...465..799L} also measured O and Mg abundances. However, O abundances are missing for some stars (\emph{e.g.} BW-f8, BW-f4) because of telluric contamination (see revision of O abundances in \citealp{2015A&A...580A..40B}), and cannot be replaced by their Mg abundances since Mg does not trace O in the galactic bulge perfectly (\emph{e.g.} \citealp{2008AJ....136..367M}). Instead, we used the $\abratio{O}{Fe}$ of \cite{2007A&A...465..799L} to build the following relation for $\abratio{\alpha}{Fe}$: for $\abratio{Fe}{H} < \SI{-0.4}{\dex}$, $\abratio{\alpha}{Fe} = \SI{0.5}{\dex}$ and for $\abratio{Fe}{H} \ge \SI{-0.4}{\dex}$, $\abratio{\alpha}{Fe} = -0.625 \times \abratio{Fe}{H}+0.25$.

\subsection{$\mu$~Leonis as a benchmark star}

\begin{figure}
  \begin{centering}
    \includegraphics[width=\columnwidth]{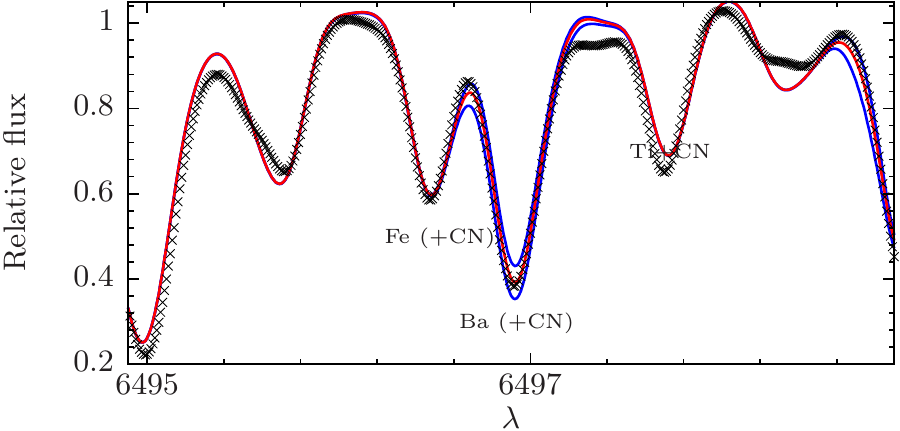}

    \includegraphics[width=\columnwidth]{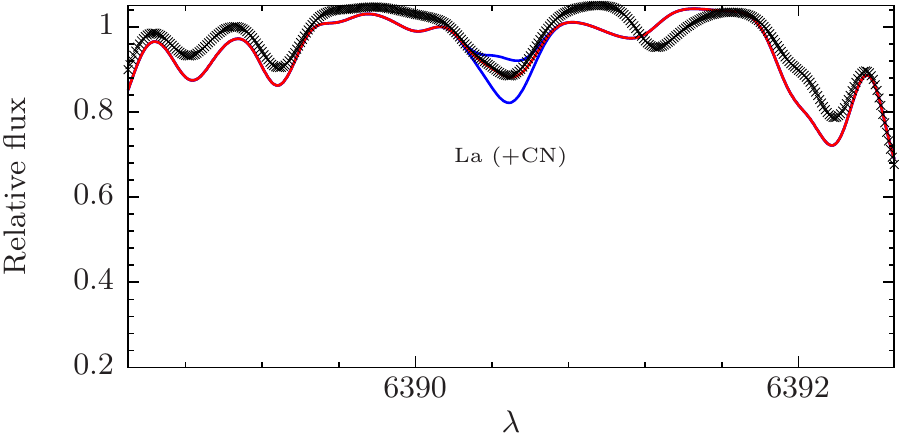}

    \includegraphics[width=\columnwidth]{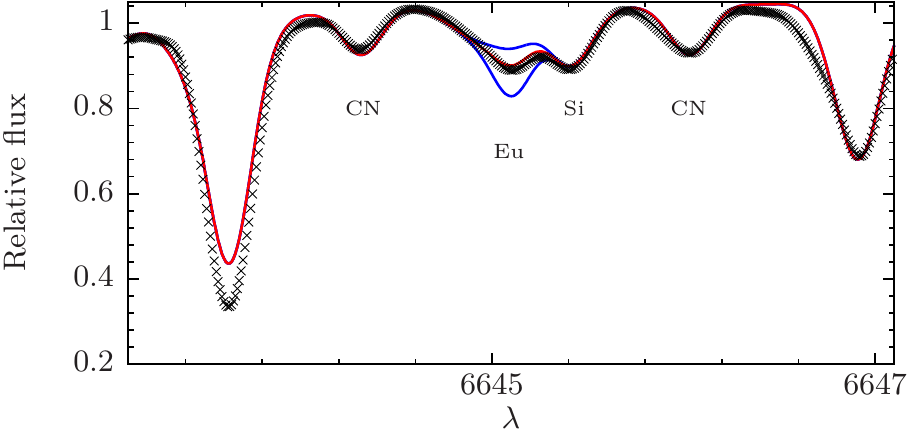}

    \caption{\label{Fig:Muleo_fits} From top to bottom: fit of \ion{Ba}{II} line at \SI{6496}{\angstrom}, \ion{La}{II} line at \SI{6390}{\angstrom}, and \ion{Eu}{II} line at \SI{6645}{\angstrom} for $\mu$ Leo (ESPaDOnS spectrum, see text for reference, degraded to UVES resolution). Black crosses: observed spectrum; red solid line: best fit; blue solid lines: spectra computed for the best-fit value of $\abratio{X}{Fe} \pm \SI{0.3}{\dex}$.}
  \end{centering}
\end{figure}

We chose the metal-rich, thin-disc giant star $\mu$~Leonis as a reference star to calibrate our line list, and, in particular, to adjust the pseudo-$\log gf$ describing the CN lines. Our benchmark analysis is based on a spectrum of $\mu$ Leo taken at CFHT with ESPaDOnS ($R \sim 80000$; see \citealp{2007A&A...465..799L}). We made the same choice of stellar parameters as \cite{2007A&A...465..799L}: $T_{\text{eff}} = \SI{4540}{\kelvin}$, $\log g = 2.3$, $[\text{M}/\text{Fe}] = \SI{0.3}{\dex}$ and $v_{\text{turb}} = \SI{1.3}{\kilo\meter\per\second}$. We also assumed the same abundance of C, N, and O as \cite{2007A&A...465..799L}: $A(\mathrm{C}= 8.85)$, $A(\mathrm{N}= 8.55)$ and $A(\mathrm{O}= 9.12)$, which translates in $\abratio{C}{Fe} = \SI{-0.01}{\dex}$, $\abratio{N}{Fe} = \SI{0.2}{\dex}$, $\abratio{O}{Fe} = \SI{-0.12}{\dex}$ (solar composition from \citealp{1998SSRv...85..161G}). Table~\ref{Tab:abundances_mu_leo} lists the individual abundances we derived for $\mu$ Leo. Figure~\ref{Fig:Muleo_fits} shows example of fits.

\begin{table*}
  \begin{center}
    \caption{\label{Tab:abundances_mu_leo} Individual abundances for $\mu$~Leonis derived from spectrum at different resolution and SNR. The column ESPaDOnS lists the abundances derived using the original ESPaDOnS spectrum. The column UVES, $\infty$ lists the abundances derived using the spectrum degraded to the UVES resolution, with the original SNR. The last two columns give the mean abundances and their standard deviation, obtained from the $101$ Monte Carlo simulations for the low and high signal-to-noise hypothesis.}
    \input{./Table.type=abundances_mu_leo.tex}
  \end{center}
\end{table*}

\subsection{Calibration of the line list}

Our selected lines are sometimes affected by a blend with atomic or molecular (CN) lines since the blends are stronger at richer metallicities. \MODIFter{However,} molecules are often poorly known and it might be necessary to determine an astrophysical $\log gf$. We sum up below a number of checks we made and the calibration we performed, when it was possible, based on high-resolution and high signal-to-noise ratio (SNR) $\mu$ Leo spectrum and the stellar parameters and CNO abundances given above.

\paragraph{Europium.}

A strong CN band popped up in the synthetic spectrum of $\mu$~Leonis at \SI{6644.5}{\angstrom} when using the original CN line lists, but this molecular band is smaller in the observed spectrum. Since this CN band modifies the blue wing of the Eu line at \SI{6645}{\angstrom}, we adjusted the corresponding pseudo-$\log gf$.

\paragraph{Barium.}

The Ba line at \SI{6496}{\angstrom} is blended with CN lines. No adjustment was needed on the CN $\log gf$. However, in our cool and metal-rich giants, the Ba line is strong (\emph{i.e.} saturated and depends weakly on the abundance), and therefore, the abundance determination suffers large errors.

\paragraph{Lanthanum.}

The La line at \SI{6262}{\angstrom} is blended with CN lines. We found that this line cannot be satisfactorily fitted in a high-resolution spectrum of the Sun, $\mu$ Leo, and Arcturus. In addition to the fact that the CN lines are poorly predicted, the shape of the La line due to the hyperfine structure never matches and is always broader than the observed line for our three reference stars. The hyperfine structure computed using the A and B coefficient of \cite{2001ApJ...556..452L}, for the $E=\SI{3250.35}{\per\centi\meter}$ and $E=\SI{19214.54}{\per\centi\meter}$ levels, is dominated by two components at $\sim\SI{6262.2}{\angstrom}$ and $\sim\SI{6262.45}{\angstrom}$, which are too broadly separated compared to the observed profile.

\begin{figure}
  \begin{centering}
    \includegraphics[height=\columnwidth,angle=-90]{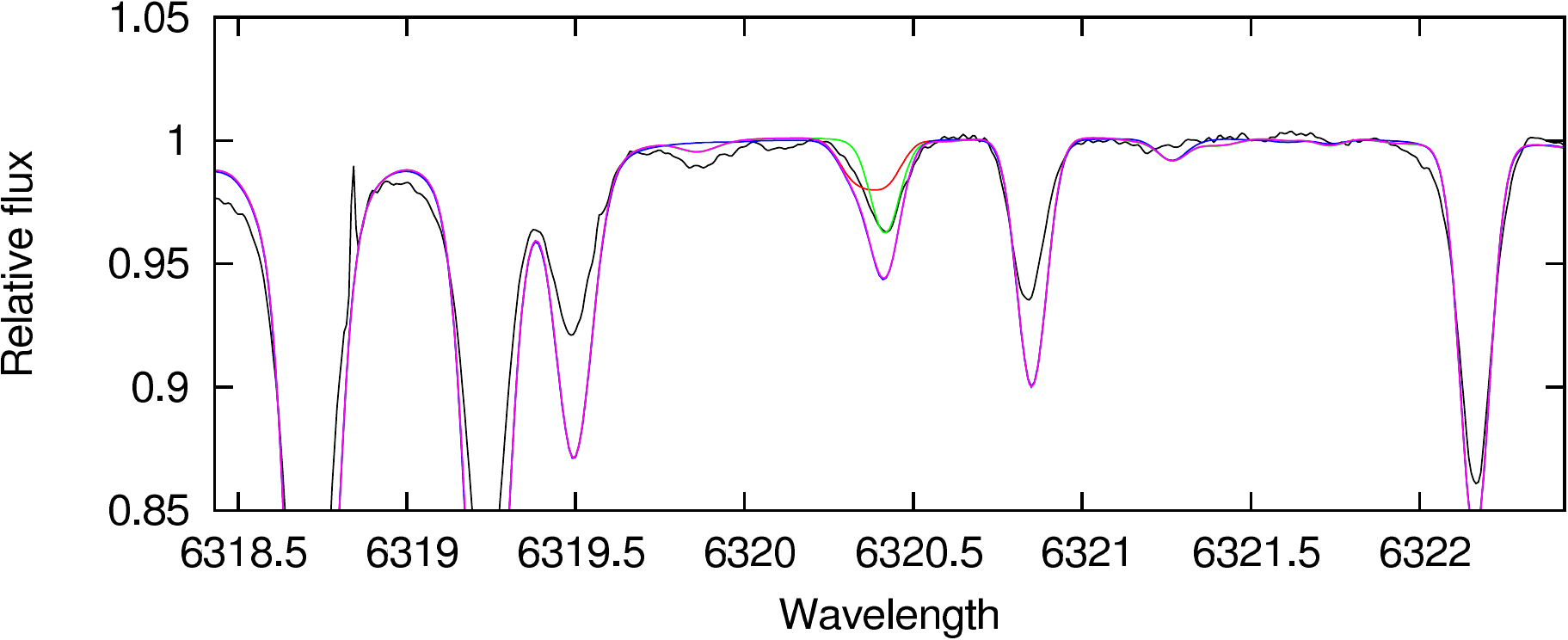}
    \caption{\label{Fig:La6320_in_Sun} Synthesis of the line at \SI{6320}{\angstrom} in the solar spectrum: observed spectrum in black \citep{1995ASPC...81...66H}, La line alone in red, Co line alone in green, total blend in purple.}
  \end{centering}
\end{figure}

Line lists predict the presence of a Co line and CN lines in the vicinity of the La line at \SI{6320}{\angstrom}. In the solar spectrum, the CN lines are completely negligible and we are left with the La and the Co lines. However, when $\abratio{La}{Fe} = \SI{0}{\dex}$ and $\abratio{Co}{Fe} = \SI{0}{\dex}$, the synthetic profile is much deeper than the observed line (Fig.~\ref{Fig:La6320_in_Sun}). One way to solve this problem is to consider this line as a pure La line and adjust the $\log gf$ accordingly, following \cite{1999A&A...351..597K}, who found $\log gf=-1.33$ instead of $-1.562$ given by VALD. We tested different hypotheses: VALD $\log gf$, considering a pure La line, considering a pure Co line or adjusted $\log gf$ so that the combination of La and Co fits the solar spectrum. However, none of the tested set of $\log gf$ allowed us to fit satisfactorily the line at \SI{6320}{\angstrom} in the spectrum of the Sun, $\mu$ Leo, and Arcturus simultaneously.

As for the La line at \SI{6390.48}{\angstrom}, the hyperfine constants are provided by \cite{2001ApJ...556..452L}. The line is contaminated by CN in its core, and some weaker CN components may appear before/after the blue/red wing of the La line (\emph{e.g.} at \SI{6389.6}{\angstrom}, \SI{6390}{\angstrom}, \SI{6391.2}{\angstrom}). As we did for the Eu line, we adjusted the $\log gf$ of the CN contaminants close to the La line to fthe Sun and $\mu$ Leo simultaneously. We did not adjust the weaker CN lines that may appear before/after the blue/red wing of the La line since their depth is comparable to the noise level in our UVES spectra as seen in the second panel of Fig.~\ref{Fig:BWRC04_fits}. We recall that Fig.~\ref{Fig:Muleo_fits} shows the fits for the high signal-to-noise ratio version of $\mu$ Leo's spectrum, that is, a better SNR than our UVES spectra.

Given the above remarks, we decided to discard the La lines at \SI{6262}{\angstrom} and \SI{6320}{\angstrom}. Our final La abundance is therefore equal to the abundance derived from the La line at \SI{6390}{\angstrom}.

\subsection{Robustness of the derived abundances}
\label{Sec:Robustness}

As a sanity check, we performed a second determination of abundances for the following lines: \ion{La}{II} line at \SI{6390}{\angstrom} and \ion{Ce}{II} line at \SI{6043}{\angstrom}. This second determination differs significantly from the main method described above by the assumptions, synthesis code, and abundance measurement.

We used the code PFANT06, described in \cite{2003A&A...404..661B} and \citet[see also \citealp{2011A&A...535A..42T}]{2005A&A...443..735C}. The heavy element line list and oscillator strengths reported in \cite{2014A&A...570A..76B} were used. Here we take the individual Mg, C, N and O abundances derived in \cite{2007A&A...465..799L} for the initial composition into account (unlike the averaged values used in Sect.~\ref{Sec:Intial_chemical_composition}). For the metal-poor stars with no oxygen derivation, B3-f2, B6-f5 and B6-f7, we assumed $\abratio{O}{Fe} = \SI{0.4}{\dex}$. Assuming a solar $\abratio{O}{Fe}$ would lead us to predict CN lines that are too strong, most obviously for the La line at \SI{6390}{\angstrom} blended with two CN A$^2$$\Pi$ - X$^2$$\Sigma$ lines ($(\nu',\nu'') = (5,1)$ P1(J=14) and $(4,0)$ P2(J=37)).

The observed spectra were smoothed with the Iraf task \emph{imfilter.gauss}, using a $\sigma = 2~\mathrm{px}$ Gaussian function. The convolution of the synthetic spectra (to take the instrument convolution, macroturbulence profile, and rotation profile into account) and the normalisation of the observed and synthetic spectra were performed locally. The CN line lists were computed by \cite{1963rspx.book.....D}. We identified the best fit to the observed spectrum with a visual inspection (unlike the $\chi^2$ algorithm used in Sect.~\ref{Sec:Methods}).

\begin{figure}
  \begin{centering}

    \includegraphics[width=0.49\columnwidth]{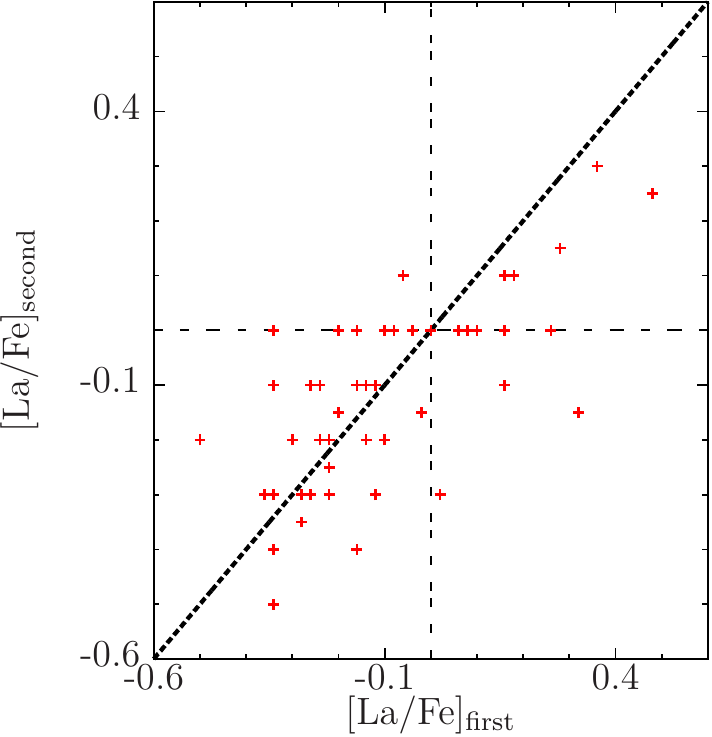}
    \includegraphics[width=0.49\columnwidth]{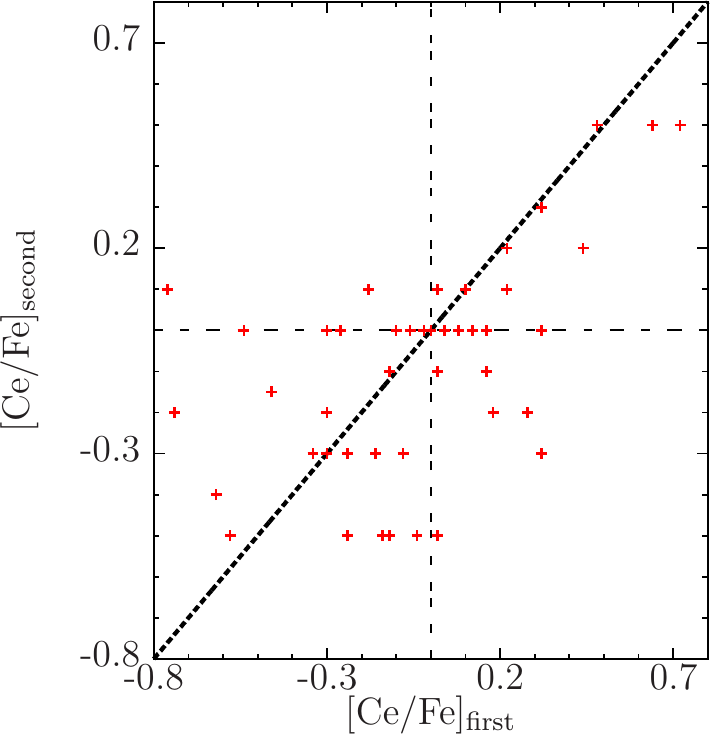}
    \caption{\label{Fig:Comparisons_methods} Comparison of abundances derived by our two methods (first method described in Sect.~\ref{Sec:Methods}, second method described in Sect.~\ref{Sec:Robustness}) for the \ion{La}{II} line at \SI{6390}{\angstrom}, \ion{Ce}{II} line at \SI{6043}{\angstrom}.}
  \end{centering}
\end{figure}

Figure~\ref{Fig:Comparisons_methods} compares the abundances obtained with the $\chi^2$ best-fit procedure (Sect.~\ref{Sec:Methods}) with those obtained with the visually accepted-fit procedure described above (Sect.~\ref{Sec:Robustness}). The mean and standard deviation of $\abratio{E}{Fe}_{\mathrm{second}} - \abratio{E}{Fe}_{\mathrm{first}}$ are \SI{-0.03}{\dex} and \SI{0.16}{\dex} for La at \SI{6390}{\angstrom} and \SI{0.03}{\dex} and \SI{0.40}{\dex} for Ce at \SI{6043}{\angstrom}, respectively. Given the differences (raw vs. smoothed observed spectrum, convolutions, normalisations, synthesis code, CN line lists, determination of the best fit), we observe an excellent agreement. The standard deviation given above is an estimator of the error on the individual abundance provided by those lines (see also Sect.~\ref{Sec:Errors}).

\subsection{Error budget}
\label{Sec:Errors}

\paragraph{Random errors.}

In order to estimate the typical (random) error bars on our individual abundances, we used our high-resolution, high SNR ESPaDOnS spectrum of $\mu$ Leo to run Monte Carlo simulations. We degraded the resolution towards the UVES resolution and we added Gaussian noise in the spectrum. We performed two SNR hypotheses ($30$ and $50$) to match the low tail and high tail of our SNR distributions (see Table~$2$ in \citealp{2007A&A...465..799L}). We created $101$ realisations for each SNR hypothesis and derived the abundances. Results are given in Table~\ref{Tab:abundances_mu_leo}. At a given SNR, when the resolution is decreased, the derived abundance tends to increase (up to $\sim \SI{+0.1}{\dex}$ in the case of Ce). At a given resolution, when the SNR decreases, the derived abundance tends to increase ($< \SI{0.1}{\dex}$). Given the change of the abundance depending on the SNR and the standard deviations provided in the last two columns of Table~\ref{Tab:abundances_mu_leo}, we consider that the conservative random errors on our derived abundances are \SI{0.15}{\dex} for Ba, La, and Eu and \SI{0.20}{\dex} for Ce and Nd in the metal rich part of our sample (to reflect their higher sensitivity to the resolution).

\paragraph{Systematic errors.}

The uncertainties in stellar parameters given in \cite{2007A&A...465..799L}, were revised in \cite{2013A&A...559A...5B}, which we adopt here, namely, $\pm \SI{150}{\kelvin}$ in temperature, $\pm 0.2$ in $\log g$ given that we used photometric gravities, $\pm \SI{0.1}{\dex}$ in $\abratio{Fe}{H}$, and $\pm \SI{0.1}{\kilo\meter\per\second}$ in microturbulence velocity. Table~\ref{Tab:Errors} reports the errors in the studied element abundances due to a change in one of the atmospheric parameters for star BWc-4. Since these parameters are covariant, the quadratic sum of individual errors represent an upper limit for the true uncertainties. The abundances are mainly sensitive to temperature and gravity.

\begin{table*}
  \begin{center}
    \caption{\label{Tab:Errors} Abundance uncertainties for star BWc-4, for uncertainties of $\Delta T_{\mathrm{eff}} = \SI{150}{\kelvin}$, $\Delta \log g = 0.2$, $\Delta v_{\mathrm{t}} = \SI{0.1}{\kilo\meter\per\second}$ and $\Delta\abratio{M}{H} = \SI{0.1}{\dex}$. $\Delta_{\pm}q$ is defined as $\abratio{X}{Fe}(q \pm \Delta q) - \abratio{X}{Fe}(q_{\text{nominal}})$, where X is the element under study and $q$ is one of the four atmospheric parameters.}
    \input{./Table.type=errors.tex}
  \end{center}
\end{table*}

\section{Results and discussion}

As explained in Sect.~\ref{Sec:Introduction}, most of the heavy elements are made by both the \emph{s}- and \emph{r}-processes of neutron capture. However few elements, like Eu, are often considered as pure \emph{r}-process elements since the \emph{r}-process is thought to over-dominate the production. For instance, the \emph{r}-process contribution to the solar Eu is \SI{94}{\percent}, according to \cite{1999ApJ...525..886A}, and \SI{97}{\percent}, according to \cite{2008ARA&A..46..241S}. On the other hand, for some elements, like Ba or La, the main contribution is thought to come from the \emph{s}-process (though the \emph{r}-process contribution is not negligible). For instance, $\approx \SI{85}{\percent}$ of the solar Ba was produced by the \emph{s}-process \citep{2000ApJ...544..302B, 2008ARA&A..46..241S}.

In Sect.~\ref{Sec:Chemical_evolution_of_the_Bulge} we compare the present results with bulge data from \cite{2012ApJ...749..175J} measured in Plaut's field at $l=\SI{0}{\degree}$, $b=\SI{-8}{\degree}$, and $l=\SI{-1}{\degree}$, $b=\SI{-8.5}{\degree}$, microlensed dwarf stars by \cite{2013A&A...549A.147B} and for stars in the metal-poor bulge globular cluster NGC $6522$ \citep{2014A&A...570A..76B}. We also plot results for NGC $6266$ \citep{2014MNRAS.439.2638Y}, which is projected towards the bulge, but is found at a rather large distance of \SI{6.7}{\kilo\parsec} from the Galactic centre, and therefore it is unclear if it is associated with the bulge component. In Sect.~\ref{Sec:Comparison_Bulge_Discs} we compare our results for the bulge with those for the Galactic thin disc and thick disc, reported by \cite{2005A&A...433..185B}, \cite{2006AJ....131..431B}, \cite{2006MNRAS.367.1329R} and \cite{2004ApJ...617.1091S}. Lastly, in Sect.~\ref{Sec:Comparison_Bulge_Discs} we compare our results to a sample of super-metal-rich stars, which were identified to be either old inner thin disc or bulge stars by \cite{2011A&A...535A..42T} and \cite{2014A&A...570A..22T}.

\subsection{Chemical evolution of the Galactic bulge}
\label{Sec:Chemical_evolution_of_the_Bulge}

\paragraph{Europium.}

\begin{figure}
  \begin{centering}
    \includegraphics[width=\columnwidth]{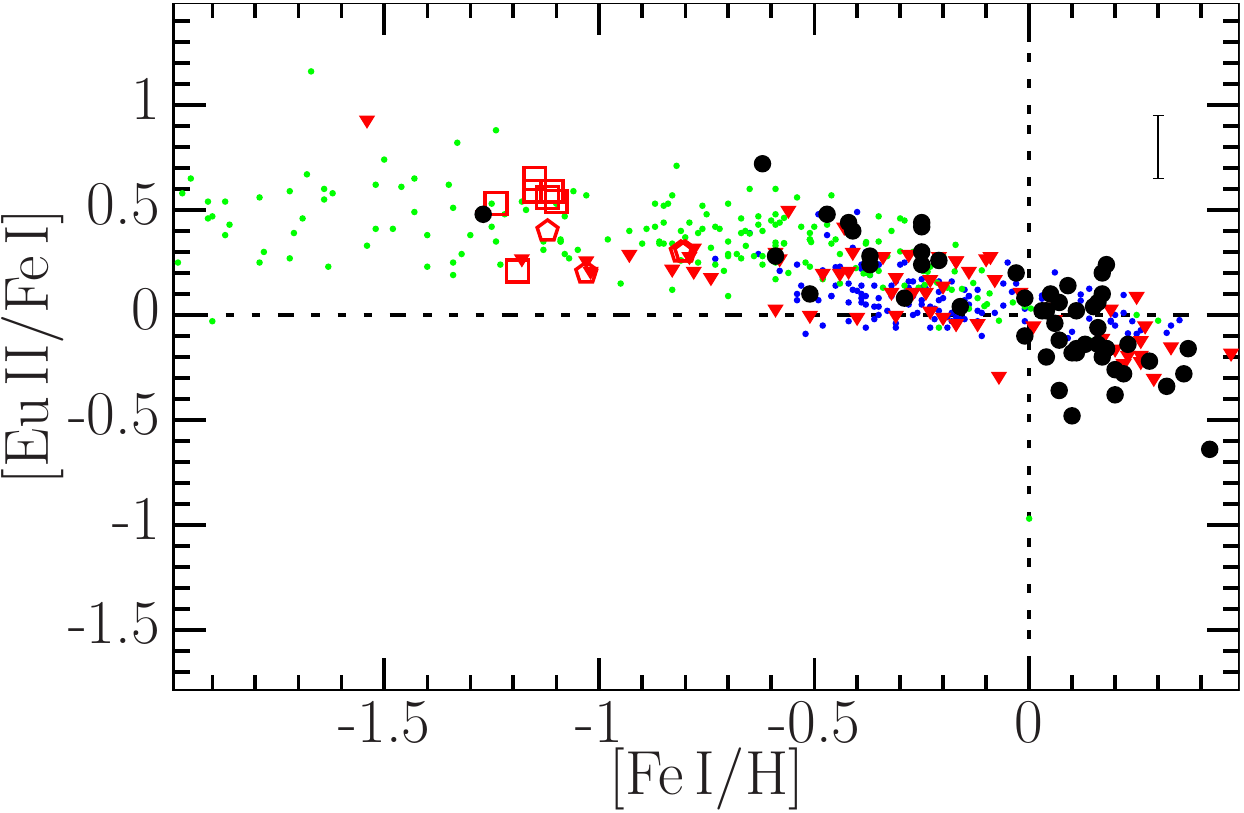}
    \caption{\label{Fig:Eu_vs_Fe} $\abratio{Eu}{Fe}$ vs. $\abratio{Fe}{H}$. Symbols: black filled circles: present work; red filled downward triangles: \cite{2012ApJ...749..175J}; red crosses: \cite{2013A&A...549A.147B}; red empty squares: M62 \citep{2014MNRAS.441.3396Y}; red empty pentagons: NGC $6522$ giants \citep{2014A&A...570A..76B}; green small dots: thick disc from \cite{2005A&A...433..185B}, \cite{2006AJ....131..431B}, \cite{2006MNRAS.367.1329R} and \cite{2004ApJ...617.1091S}; blue small dots: thin disc from \cite{2005A&A...433..185B} and \cite{2006MNRAS.367.1329R}.}
  \end{centering}
\end{figure}

Figure~\ref{Fig:Eu_vs_Fe} shows our bulge trend for $\abratio{Eu}{Fe}$. For $\abratio{Fe}{H} \geq \SI{-0.5}{\dex}$, we clearly observe a decreasing $\abratio{Eu}{Fe}$ with increasing $\abratio{Fe}{H}$. While our sample contains only a couple of stars with $\abratio{Fe}{H} \leq \SI{-0.5}{\dex}$, their abundance lie on a plateau of enhanced $\abratio{Eu}{Fe}$ in agreement with the constant Eu abundances derived in metal-poor bulge globular clusters (M22 as of \citealp{2014MNRAS.439.2638Y} or NGC6522, \citealp{2014A&A...570A..76B}). At higher metallicities ($\abratio{Fe}{H} > \SI{-0.5}{\dex}$), our bulge $\abratio{Eu}{Fe}$ clearly decreases with increasing metallicity in agreement with the behaviour also observed by \cite{2012ApJ...749..175J}. Thus, we found for Eu the expected $\alpha$-like behaviour for an \emph{r}-element produced by massive stars (\emph{e.g.} Fig.~$14$ in \citealp{2008ARA&A..46..241S}); the decreasing trend reflects the efficient production of iron by SN~Ia after time delay between a few \SI{100}{\mega\Year} to $\sim \SI{1}{\giga\Year}$ \citep{2007A&A...467..123B}.

\paragraph{Barium, lanthanum, cerium and neodymium.}

\begin{figure}
  \begin{centering}
    \includegraphics[width=\columnwidth]{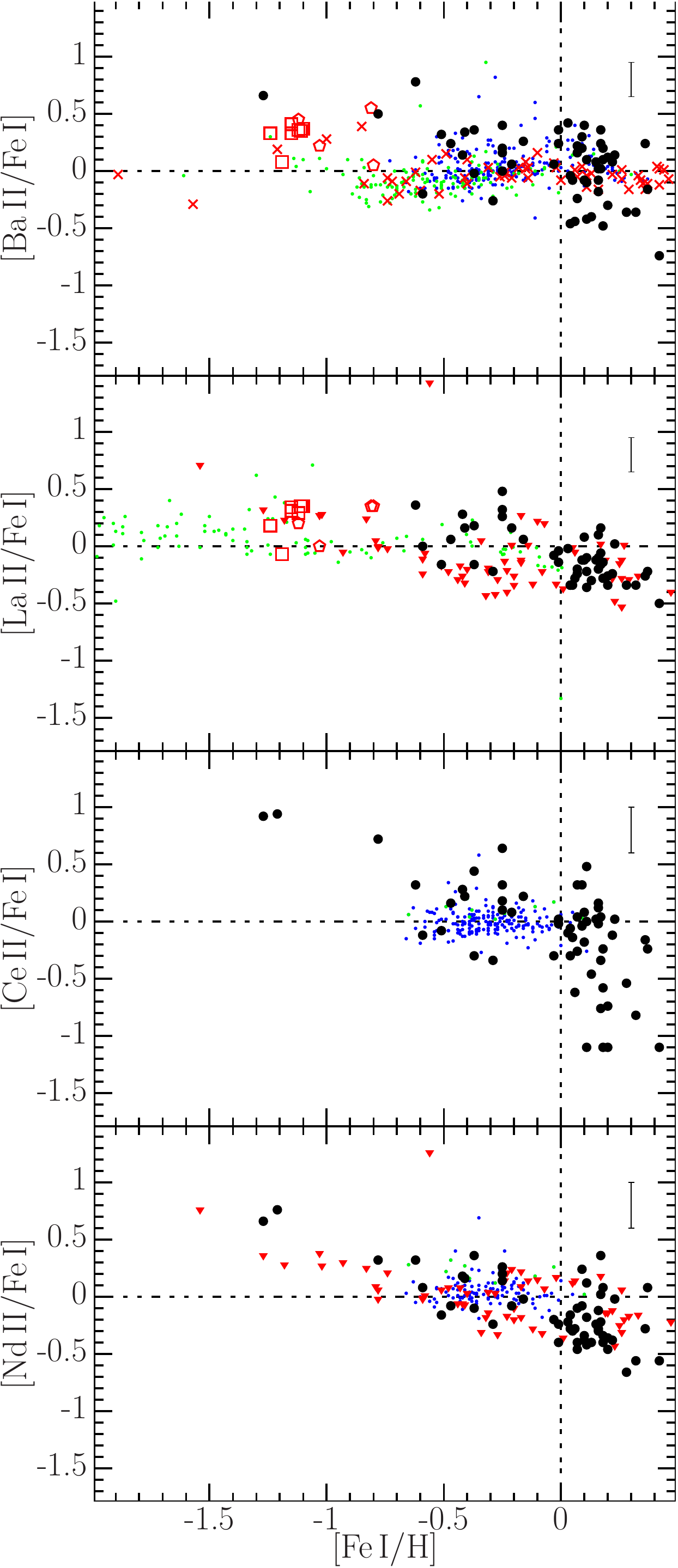}
    \caption{\label{Fig:Heavy_s_elements} $\abratio{Ba, La, Ce, Nd}{Fe}$ vs. $\abratio{Fe}{H}$. Same symbols as in Fig.~\ref{Fig:Eu_vs_Fe}.}
  \end{centering}
\end{figure}

Figure~\ref{Fig:Heavy_s_elements} shows our bulge trends for Ba, La, Ce, and Nd, all belonging to the second peak of the \emph{s}-process. We find that $\abratio{Ba}{Fe}$, $\abratio{La}{Fe}$, $\abratio{Ce}{Fe}$ and $\abratio{Nd}{Fe}$ decrease slightly with increasing metallicities. We obtain a very good agreement between our trends for $\abratio{La}{Fe}$ and $\abratio{Nd}{Fe}$ and those by \cite{2012ApJ...749..175J}. Our most metal-poor stars have a $\abratio{Ba}{Fe}$ and $\abratio{La}{Fe}$ compatible with results for the globular clusters NGC $6522$ \citep{2014A&A...570A..76B} and NGC $6266$ \cite{2014MNRAS.439.2638Y}.

\begin{figure}
  \begin{centering}
    \includegraphics[width=\columnwidth]{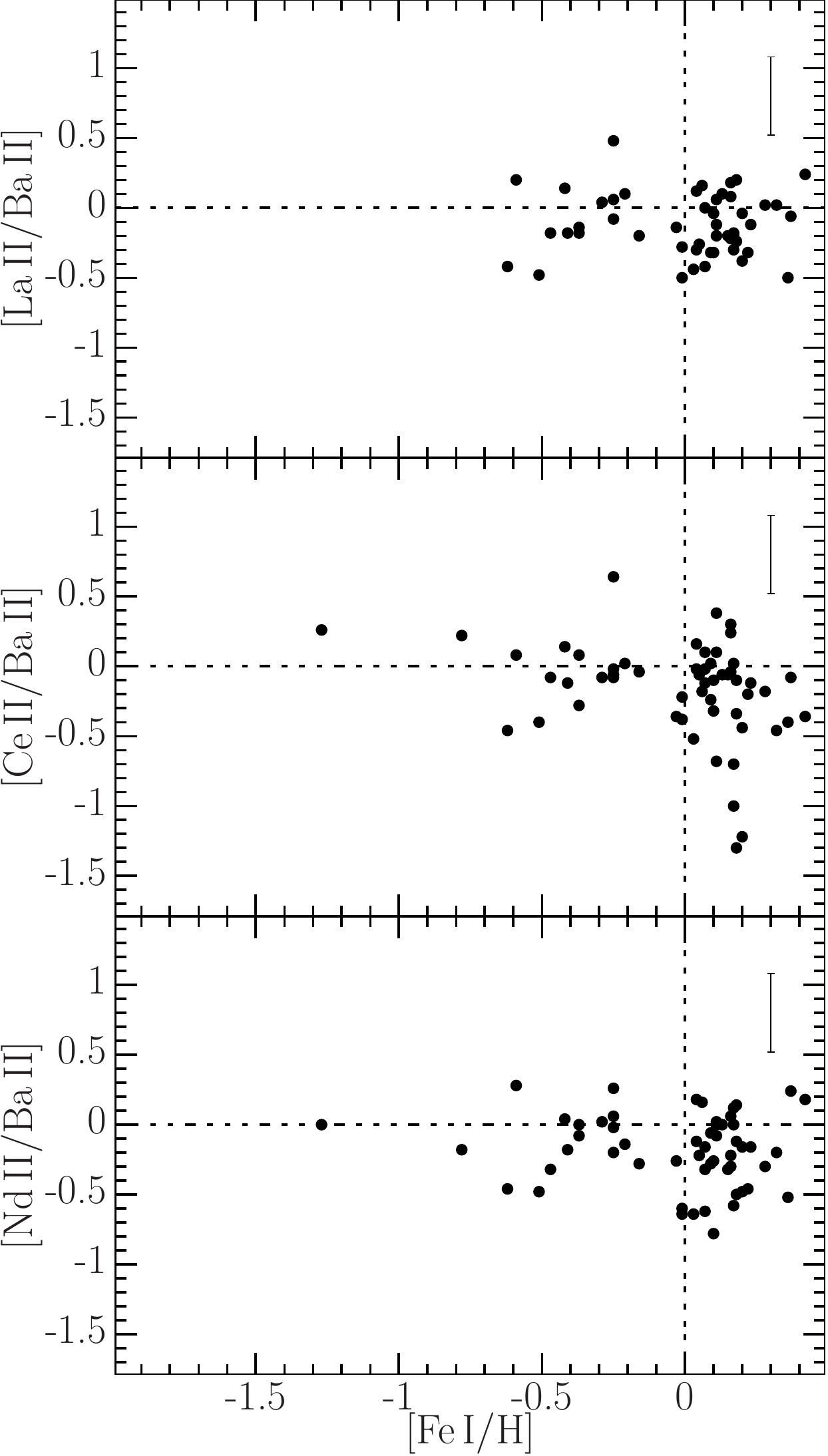}
    \caption{\label{Fig:Heavy_to_heavy_1} $\abratio{La, Ce, Nd}{Ba}$ vs. $\abratio{Fe}{H}$. Black dots: our bulge sample.}
  \end{centering}
\end{figure}

\begin{figure}
  \begin{centering}
    \includegraphics[width=\columnwidth]{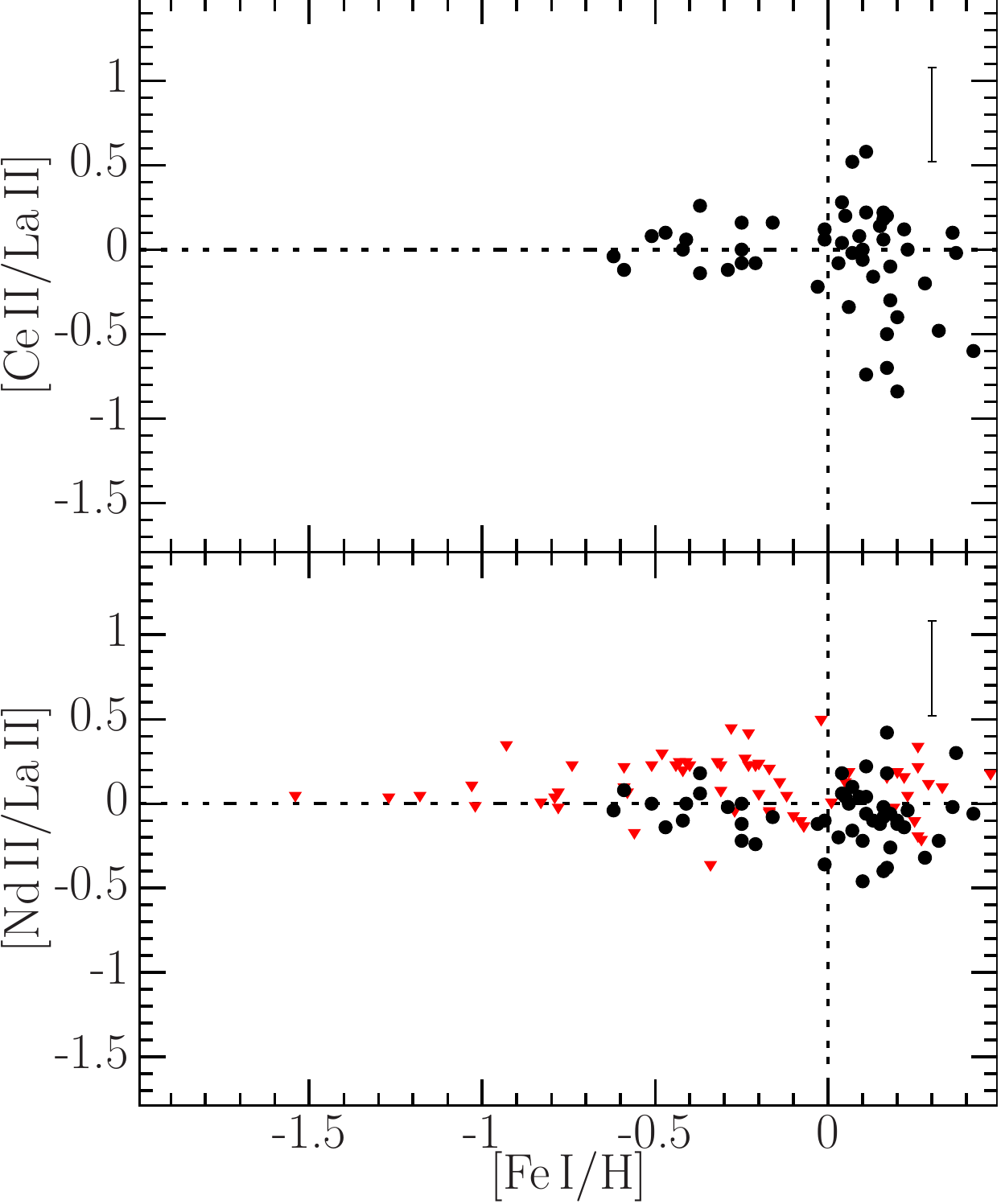}
    \caption{\label{Fig:Heavy_to_heavy_2} $\abratio{Ce, Nd}{La}$ vs. $\abratio{Fe}{H}$. Same symbols as in Fig.~\ref{Fig:Eu_vs_Fe}.}
  \end{centering}
\end{figure}

Ratios among neutron-capture elements can help to diagnose nucleosynthesis processes (\emph{e.g.} \citealp{2013ApJ...771...67I}). Figures~\ref{Fig:Heavy_to_heavy_1} and \ref{Fig:Heavy_to_heavy_2} show $\abratio{La, Ce, Nd}{Ba}$ vs. $\abratio{Fe}{H}$ and $\abratio{Ce, Nd}{La}$ vs. $\abratio{Fe}{H}$ respectively. Those ratios are around the solar value for all metallicities, therefore, we can conclude that these elements are produced in the same stars or at least in the same proportions in any event.

\begin{figure}
  \begin{centering}
    \includegraphics[width=\columnwidth]{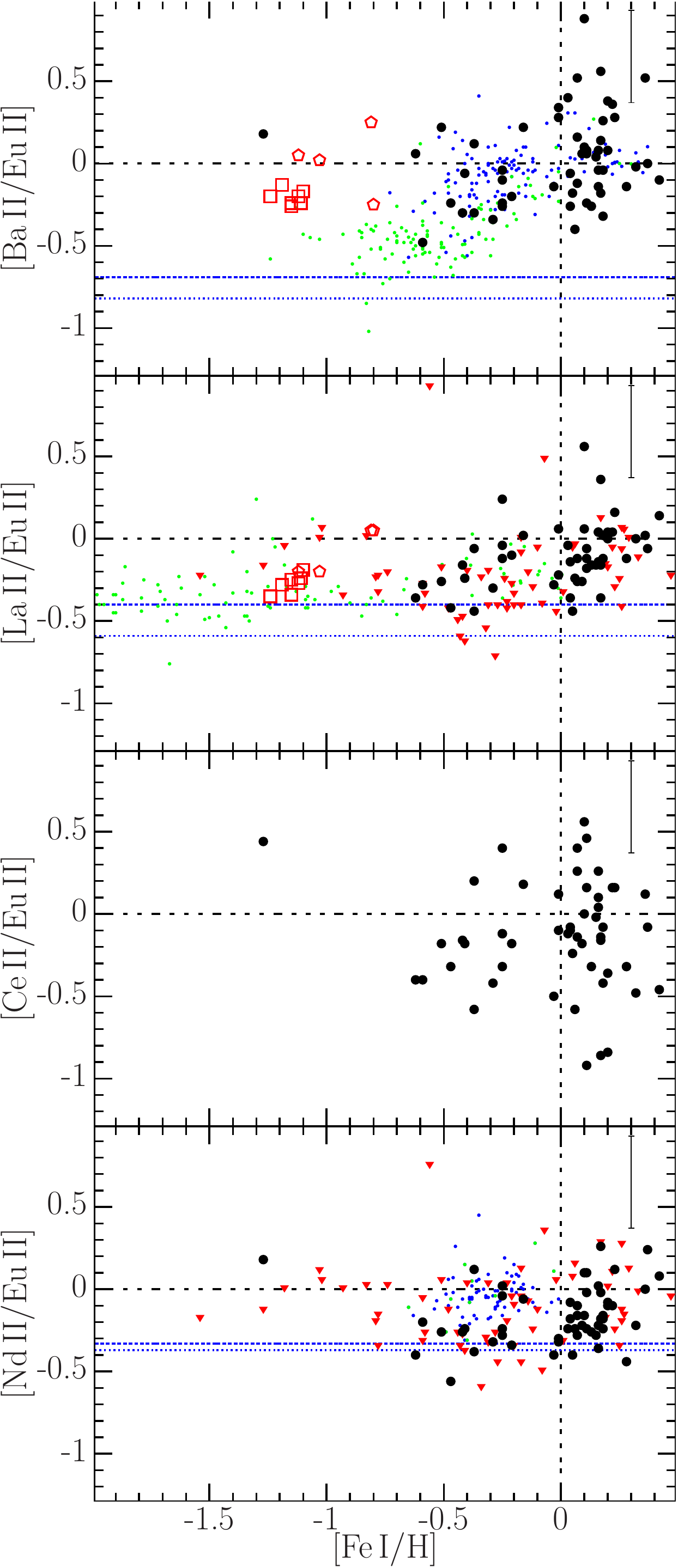}
    \caption{\label{Fig:s_to_r_process} $\abratio{Ba, La, Ce, Nd}{Eu}$ vs. $\abratio{Fe}{H}$. Same symbols as in Fig.~\ref{Fig:Eu_vs_Fe}. Dotted and dashed blue lines: pure \emph{r}-process values provided by \cite{1999ApJ...525..886A} and \cite{2008ARA&A..46..241S}, respectively.}
  \end{centering}
\end{figure}

In order to disentangle the contribution of the \emph{r}- and \emph{s}-process, it is useful to plot the abundances among the heavy elements, as for example ratios to a reference \emph{s}-element, such as Ba or La, and a reference \emph{r}-element, such as Eu. Figure~\ref{Fig:s_to_r_process} shows the trends for $\abratio{Ba, La, Ce, Nd}{Eu}$ vs. $\abratio{Fe}{H}$. These trends show essentially that the dominantly \emph{s}-elements Ba, La, Ce, and Nd are underabundant relative to the \emph{r}-element Eu, and the ratio is below solar at the lower metallicities where $\abratio{Eu}{Fe}$ is overabundant \MODIFter{(Fig.~\ref{Fig:Eu_vs_Fe})}. The plots of $\abratio{La}{Eu}$ and $\abratio{Nd}{Eu}$ are better defined compared to those for Ba and Ce; see Sect.~\ref{Sec:Abundances}. We find a good agreement between our $\abratio{La, Nd}{Eu}$ trends and those of \cite{2012ApJ...749..175J}, as well as with the data of \cite{2014MNRAS.439.2638Y} for the metal-poor cluster M$62$. Our bulge metal-poor stars ($\SI{-0.6}{\dex} < \abratio{Fe}{H} < \SI{0}{\dex}$) have mean $\abratio{La, Nd}{Eu}$ ratios comparable, within the uncertainties, to the value of the pure \emph{r}-process ratio: for La, \SI{-0.17}{\dex} (std. dev: \SI{0.17}{\dex}) vs. $\abratio{La_{\mathrm{r}}}{Eu_{\mathrm{r}}} ~ \SI{-0.4}{\dex}$ \citep{1999ApJ...525..886A}; for Nd, \SI{-0.24}{\dex} (std. dev: \SI{0.16}{\dex}) vs. $\abratio{Nd_{\mathrm{r}}}{Eu_{\mathrm{r}}} ~ \SI{-0.35}{\dex}$ \citep{1999ApJ...525..886A, 2008ARA&A..46..241S}. On the other hand, the super-solar part of our sample exhibit higher ratios: for La, \SI{-0.06}{\dex} (std. dev: \SI{0.20}{\dex}); for Nd, \SI{-0.12}{\dex} (std. dev: \SI{0.17}{\dex}). This points to an increasing trend of $\abratio{La, Nd}{Eu}$ from \SI{-0.6}{\dex} to \SI{+0.4}{\dex}, that is, the rise of the \emph{s}-process in the chemical enrichment history of the Galactic bulge. The Ba shows a similar behaviour, that is, the distribution is compatible with an increasing trend towards super-solar metallicities, but never reaches the pure \emph{r}-process level. Also, the two most metal-poor stars (around \SI{-1.3}{\dex}) exhibit very high $\abratio{Ba, Nd}{Eu}$. Possibilities to explain such overabundances are discussed in \cite{2014A&A...570A..76B}, \cite{2011Natur.472..454C}. It is also interesting to see that $\abratio{La, Nd}{Eu}$ ratios increase moderately with increasing metallicity for supersolar metallicities. There is a gap in metallicity at $-0.2 < \abratio{Fe}{H} < \SI{0}{\dex}$, and we cannot check where this increase in La, Nd starts. This finding, together with our previous finding of low (subsolar) $\abratio{La, Nd, Ce}{Fe}$ values in our sample, supports a scenario of a weak contribution of the \emph{s}-process in the production of the second-peak elements.

\subsection{Comparison between our sample vs. Galactic discs and bulge-like solar-neighbourhood stars}
\label{Sec:Comparison_Bulge_Discs}

Figure~\ref{Fig:Eu_vs_Fe} shows good agreement between our bulge $\abratio{Eu}{Fe}$ trend and the trends for the Galactic thick and thin discs derived by \cite{2005A&A...433..185B}, \cite{2006AJ....131..431B}, \cite{2006MNRAS.367.1329R}, and \cite{2004ApJ...617.1091S}. It is not surprising, given the fact that Eu is processed by massive stars, and thus, is expelled in the interstellar medium at early stages of the galactic evolution.

\begin{table}
  \begin{center}
    \caption{\label{Table:s_elements} Mean abundances and their associated standard deviation for our bulge Ba, La, Nd, Ce distributions.}
    \input{./Table.type=s_elements.tex}
  \end{center}
\end{table}

If one compares the abundance diagrams of second peak \emph{s}-elements obtained for the different galactic components (Fig.~\ref{Fig:Heavy_s_elements}), we find that $\abratio{Ba, La, Ce, Nd}{Fe}$ ratios in bulge stars tend to lie above \SI{0}{\dex} for $\abratio{Fe}{H} < -0.2 \mathrm{\,-\,} \SI{0}{\dex}$ and below \SI{0}{\dex} for $\abratio{Fe}{H} > \SI{0}{\dex}$ (Table~\ref{Table:s_elements}), while the discs show a flatter distribution at all metallicities. Though the comparison between MW discs and bulge cannot be repeated, the full metallicity range of our bulge sample (paucity of data points above $\sim \SI{0.1}{\dex}$ for the MW discs), our findings may suggest a decreasing trend for $\abratio{Ba, La, Ce, Nd}{Fe}$ in the bulge stars. We emphasize that $\mu$ Leo ($\abratio{Fe}{H} = \SI{0.3}{\dex}$) behaves like the metal-rich part of our sample: in particular, $\mu$ Leo has $\abratio{La}{Fe} = \SI{-0.36}{\dex}$ and $\abratio{Nd}{Fe} = \SI{-0.7}{\dex,}$ while our sample reaches $\abratio{La}{Fe} = \SI{-0.35}{\dex}$ and $\abratio{Nd}{Fe} = \SI{-0.6}{\dex}$ at similar metallicity. Systematic effects between dwarf and giant stars may explain the negative slope observed for our giant bulge stars. Assuming that the slope indeed has astrophysical origin, and since Ba, La, Ce, Nd are mainly produced by the \emph{s}-process and Fe is mainly produced by SN~Ia, the decrease of $\abratio{Ba, La, Ce, Nd}{Fe}$ with increasing metallicities could be due to a weak production of Ba, La, Ce, Nd by the \emph{s}-process and/or a strong production of Fe by SN~Ia. A very active production of iron by SN~Ia is unlikely since it would affect other $\abratio{X}{Fe}$ vs. $\abratio{Fe}{H}$ diagrams (\emph{e.g.} $\alpha$-elements) and it would require a higher rate of binary system formation in the bulge than in the MW discs. On the other hand, it is more reasonable to assume a weaker contribution of the \emph{s}-process, compared to the Galactic thin and thick discs, and therefore, assume that the second peak elements would be mainly produced by the \emph{r}-process. The increase of $\abratio{Ba, La, Ce, Nd}{Eu}$ vs. $\abratio{Fe}{H}$ may indicate that the \emph{s}-process from AGB stars starts to operate at a metallicity around solar (Fig.\ref{Fig:s_to_r_process}), that is, at a later epoch in the bulge history compared to the discs.

\begin{figure}
  \begin{centering}
    \includegraphics[width=\columnwidth]{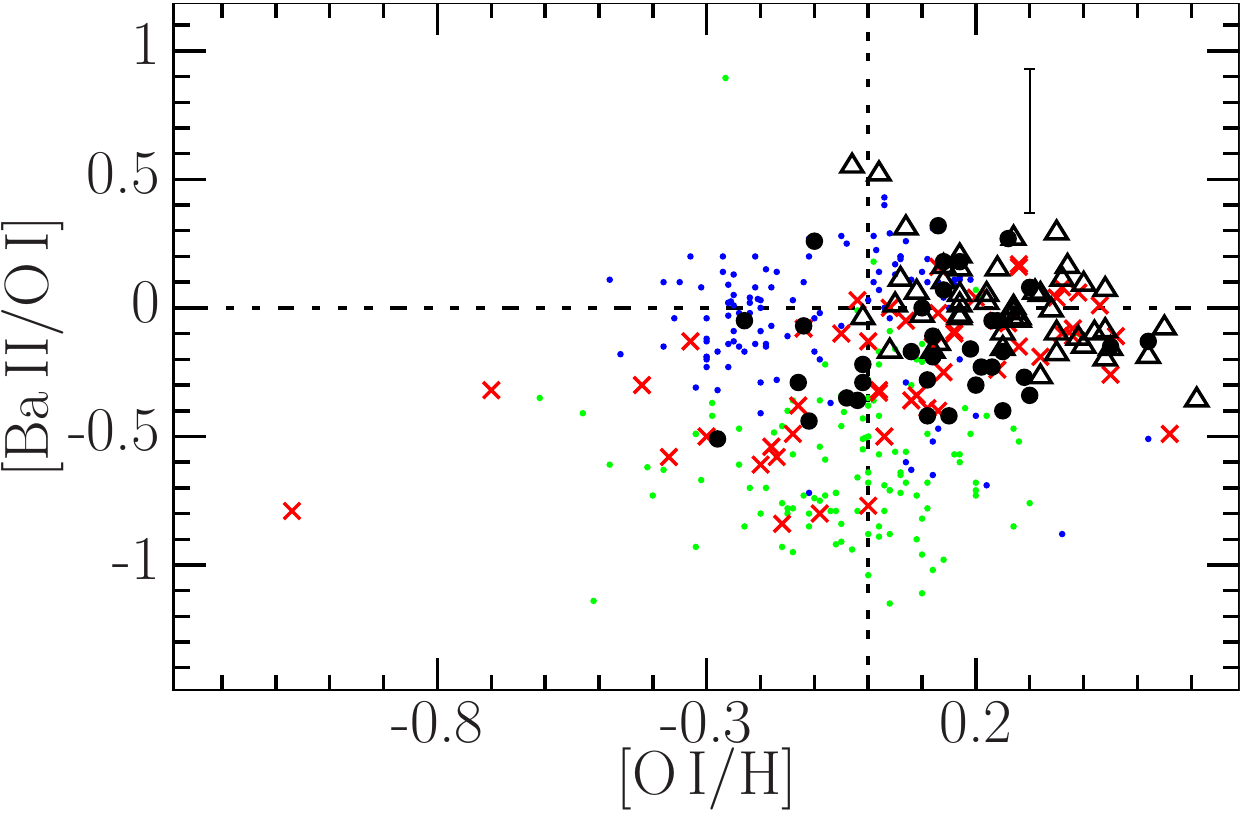}
    \caption{\label{Fig:BaO_vs_O} $\abratio{Ba}{O}$ vs. $\abratio{O}{H}$. Same symbols as in Fig.~\ref{Fig:Eu_vs_Fe} and black empty triangles: super metal-rich, bulge-like \cite{2014A&A...570A..22T}.}
  \end{centering}
\end{figure}

In order to better disentangle the Galactic populations, we plot $\abratio{Ba}{O}$ vs. $\abratio{Fe}{H}$ in Fig.~\ref{Fig:BaO_vs_O}. This figure shows that our $\abratio{Ba}{O}$ results overlap well the thick disc distribution from \cite{2005A&A...433..185B} and \cite{2006MNRAS.367.1329R}. Finally, all these stellar populations are clearly distinct from the thin disc in the metal poor regime, that is, where the thin and thick discs themselves are chemically distinct. On the other hand, the $\abratio{Ba}{O}$ distribution derived by \cite{2014A&A...570A..22T} for bulge-like solar-neighbourhood stars is more similar to the thin disc distribution than our bulge distribution. This \MODIFter{similarity} reinforces the finding by \cite{2011A&A...535A..42T} and \cite{2014A&A...570A..22T} that this sample should rather be associated with an old inner thin disc rather than a bulge-like population.

\section{Conclusions}

The heavy elements appear to be a useful indicator of chemical enrichment in different environments. We find evidence for differences between their abundances in the thin and thick discs and bulge stars. We emphasize that the Fornax dSph or the Large Magellanic Cloud have their $\abratio{Ba,La}{Fe}$ ratios \emph{increasing} with increasing metallicities (e.g \citealp{2009ARA&A..47..371T}, \citealp{2013ApJ...778..149M}, \citealp{2013A&A...560A..44V}), making evident that the chemical enrichment of these galaxies is dominated by the s-process and the trends are clearly distinct from those of stellar populations in the Galaxy.

We have derived abundances of the neutron-capture elements Ba, La, Ce, Nd, and Eu in \num{56} bulge red giants. Their metallicities are in the range $-1.3 < \abratio{Fe}{H} < \SI{0.3}{\dex}$, which, if one follows the multiple populations proposed by \cite{2010A&A...519A..77B}, \cite{2011A&A...534A..80H}, \cite{2013MNRAS.430..836N}, or \cite{2014A&A...569A.103R}, would mean that our sample includes stars in two stellar populations: old metal-poor with isotropic kinematics and metal-rich stars with bar-like kinematics. Even so, the behaviour of heavy-element-to-iron abundances vs. metallicity show some clear trends, among which we can cite as more important:

The dominantly \emph{r}-element Eu shows a $\abratio{Eu}{Fe}$ vs. $\abratio{Fe}{H}$ ratio similar to that of alpha-elements (Fig.~\ref{Fig:Eu_vs_Fe}), that is, a flattening of the distribution towards low metallicities ($\abratio{Fe}{H} < \sim \SI{-0.5}{\dex}$), which is compatible with a plateau. \MODIFter{The trend $\abratio{Eu}{Fe}$ vs. $\abratio{Fe}{H}$} also reveals a turnover at around $\abratio{Fe}{H} \sim \SI{-0.6}{\dex}$ due to Fe enrichment from type Ia SNe. It is interesting to compare the metallicity of the turnover with that for bulge microlensed stars by \citet[their Fig.~$25$]{2013A&A...549A.147B} located at around $\abratio{Fe}{H} \sim \SI{-0.7}{\dex}$, and $\abratio{Fe}{H} \sim \SI{-0.4}{\dex}$ for thick disc stars, which can be seen in Fig.~$9$ of \cite{2004A&A...415..155B}. This could indicate a small difference in the chemical enrichment of the bulge and thick disc (different SN~II/SN~Ia ratios along time). For $\abratio{Fe}{H} > \SI{-0.6}{\dex}$, there is a decreasing $\abratio{Eu}{Fe}$ vs. $\abratio{Fe}{H}$, reaching the solar value at the solar metallicity, and decreasing further for higher metallicities. On the other hand, in Fig.~\ref{Fig:s_to_r_process}, the $\abratio{La}{Eu}$ and $\abratio{Nd}{Eu}$ seem to start increasing at around $\abratio{Fe}{H} \sim \SI{-0.4}{\dex}$ due to a triggering of \emph{s}-element enrichment by AGB stars.

Figure~\ref{Fig:Heavy_s_elements} provide trends for $\abratio{Ba}{Fe}$, $\abratio{La}{Fe}$, $\abratio{Ce}{Fe}$ and $\abratio{Nd}{Fe}$ vs. $\abratio{Fe}{H}$, showing that they decrease steadily with increasing metallicities. One possible interpretation would be that the second peak elements are mainly produced by the \emph{r}-process and that the decrease with metallicity is due to Fe enrichment by SN~Ia. However, the increase of the $\abratio{La}{Eu}$ and $\abratio{Nd}{Eu}$ ratios with increasing metallicity for supersolar metallicities is due to \emph{s}-element enrichment by AGB stars. Consequently, in that metallicity regime, the decrease in abundance (relative to iron) of the second peak neutron capture elements must be due to the \emph{s}-process production not being quite strong enough to counterbalance the iron production by SN~Ia.

For metal-poor stars in the range $\SI{-0.6}{\dex} < \abratio{Fe}{H} < \SI{-0.3}{\dex}$, $\abratio{La, Nd}{Eu}$ ratios are located close to the pure \emph{r}-process expected value (\citealp{1999ApJ...525..886A}; \citealp{2008ARA&A..46..241S}), indicating that for these older stars all these elements are produced by the \emph{r}-process.

Finally, Fig.~\ref{Fig:BaO_vs_O} showing $\abratio{Ba}{O}$ vs. $\abratio{O}{H}$ shows that our bulge stars and the thick disc data from \cite{2005A&A...433..185B} and \cite{2006MNRAS.367.1329R} are compatible. The same figure shows that all these stellar populations are clearly distinct from the thin disc. Another interesting product of Fig.~\ref{Fig:BaO_vs_O} is the comparison with metal-rich stars by \cite{2011A&A...535A..42T} and \cite{2014A&A...570A..22T}, indicating that their sample better resembles an old inner thin disc than a bulge-like population.

\begin{acknowledgements}
We thank the anonymous referee for his/her remarks, which resulted in an improved discussion of the results. MVDS acknowledges a FAPESP post-doctoral fellowship n$^{\circ}$ 2013/50361-0. BB acknowledges partial financial support by CNPq and FAPESP. VH acknowledges support of the French Agence Nationale de la Recherche under contract ANR-2010-BLAN- 0508-01OTP. MZ and DM are supported by Fondecyt Regular 1110393 and 1130196, by the BASAL Center for Astrophysics and Associated Technologies CATA PFB-06, the FONDAP Center for Astrophysics 15010003, by the Chilean Ministry for the Economy, Development, and Tourism Programa Iniciativa Cient\'{\i}fica Milenio through grant IC120009, and awarded to the Millennium Institute of Astrophysics, MAS.
\end{acknowledgements}

\bibliographystyle{aa}
\bibliography{Bibliography}

\end{document}

%% file: Table.type=abundances_bulge_stars.tex
\begin{tabular}{lS[table-format=2.2]S[table-format=2.2]S[table-format=2.2]S[table-format=2.2]S[table-format=2.2]S[table-format=2.2]}
  \hline
  \hline
  \noalign{\smallskip}
  Stars   & {$\abratio{Fe}{H}$}         & {$\abratio{Ba}{Fe}$} & {$\abratio{La}{Fe}$} & {$\abratio{Ce}{Fe}$} & {$\abratio{Nd}{Fe}$} & {$\abratio{Eu}{Fe}$}\\
  \noalign{\smallskip}
          &                             & {6496} & {6390} & {6043} & {6740} & {6645} \\
  \cline{2-7}
  \noalign{\smallskip}
          & \multicolumn{6}{c}{\SI{}{\dex}}       \\
  \hline
  \noalign{\smallskip}
B3-b1	&	-0.78	& 	 0.50	&	  {NaN}	&	 0.72	&	 0.32	&	  {NaN} \\
B3-b2	&	 0.18	& 	 0.20	&	  {NaN}	&	{NaN}	&	 0.08	&	 0.24 \\
B3-b3	&	 0.18	& 	-0.48	&	-0.28	&	-0.58	&	-0.34	&	-0.16 \\
B3-b4	&	 0.17	& 	 0.02	&	-0.16	&	 0.04	&	 0.02	&	 0.20 \\
B3-b5	&	 0.11	& 	-0.42	&	-0.36	&	{NaN}	&	-0.42	&	-0.18 \\
B3-b7	&	 0.20	& 	 0.12	&	-0.26	&	{NaN}	&	-0.36	&	-0.26 \\
B3-b8	&	-0.62	& 	 0.78	&	 0.36	&	 0.32	&	 0.32	&	 0.72 \\
B3-f1	&	 0.04	& 	-0.04	&	-0.34	&	-0.06	&	-0.16	&	 0.02 \\
B3-f2	&	-0.25	& 	 0.00	&	 0.48	&	 0.64	&	 0.26	&	 0.24 \\
B3-f3	&	 0.06	& 	-0.44	&	-0.28	&	-0.62	&	-0.28	&	-0.04 \\
B3-f4	&	 0.09	& 	 0.30	&	  {NaN}	&	 0.32	&	 0.24	&	  {NaN} \\
B3-f5	&	 0.16	& 	-0.08	&	 0.10	&	 0.16	&	-0.30	&	 0.06 \\
B3-f7	&	 0.16	& 	-0.18	&	-0.10	&	 0.12	&	-0.12	&	-0.14 \\
B3-f8	&	 0.20	& 	-0.30	&	-0.34	&	-0.74	&	-0.46	&	-0.38 \\
B6-b1	&	 0.07	& 	-0.24	&	-0.24	&	-0.26	&	-0.40	&	-0.12 \\
B6-b2	&	-0.01	& 	 0.24	&	-0.04	&	 0.02	&	-0.40	&	-0.10 \\
B6-b3	&	 0.10	& 	-0.08	&	-0.12	&	-0.18	&	-0.34	&	-0.18 \\
B6-b4	&	-0.41	& 	 0.34	&	 0.16	&	 0.22	&	 0.16	&	 0.40 \\
B6-b5	&	-0.37	& 	-0.02	&	-0.16	&	-0.30	&	-0.10	&	 0.28 \\
B6-b6	&	 0.11	& 	-0.10	&	-0.22	&	 0.00	&	-0.18	&	-0.16 \\
B6-b8	&	 0.03	& 	 0.42	&	-0.02	&	-0.10	&	-0.22	&	 0.02 \\
B6-f1	&	-0.01	& 	 0.36	&	-0.14	&	-0.02	&	-0.24	&	 0.08 \\
B6-f2	&	-0.51	& 	 0.32	&	-0.16	&	-0.08	&	-0.16	&	 0.10 \\
B6-f3	&	-0.29	& 	-0.26	&	-0.22	&	-0.34	&	-0.24	&	 0.08 \\
B6-f5	&	-0.37	& 	 0.36	&	 0.18	&	 0.44	&	 0.36	&	 0.24 \\
B6-f7	&	-0.42	& 	 0.14	&	 0.28	&	 0.28	&	 0.18	&	 0.44 \\
B6-f8	&	 0.04	& 	-0.46	&	-0.34	&	-0.30	&	-0.28	&	-0.20 \\
BL-01	&	-0.16	& 	 0.26	&	 0.06	&	 0.22	&	-0.02	&	 0.04 \\
BL-03	&	-0.03	& 	 0.06	&	-0.08	&	-0.30	&	-0.20	&	 0.20 \\
BL-04	&	 0.13	& 	-0.40	&	-0.30	&	-0.46	&	-0.40	&	-0.14 \\
BL-05	&	 0.16	& 	 0.02	&	-0.20	&	-0.02	&	-0.28	&	-0.06 \\
BL-07	&	-0.47	& 	 0.24	&	 0.06	&	 0.16	&	-0.08	&	 0.48 \\
BW-b2	&	 0.22	& 	 0.08	&	-0.24	&	-0.12	&	-0.38	&	-0.28 \\
BW-b4	&	 0.07	& 	 0.16	&	  {NaN}	&	 0.04	&	-0.46	&	-0.36 \\
BW-b5	&	 0.17	& 	 0.36	&	 0.16	&	-0.34	&	-0.22	&	-0.20 \\
BW-b6	&	-0.25	& 	 0.16	&	  {NaN}	&	 0.10	&	 0.14	&	 0.42 \\
BW-b7	&	 0.10	& 	 0.40	&	 0.08	&	 0.08	&	-0.38	&	-0.48 \\
BW-f1	&	 0.32	& 	-0.36	&	-0.34	&	-0.82	&	-0.56	&	-0.34 \\
BW-f4	&	-1.21	& 	  {NaN}	&	  {NaN}	&	 0.94	&	 0.76	&	  {NaN} \\
BW-f5	&	-0.59	& 	-0.20	&	 0.00	&	-0.12	&	 0.08	&	 0.28 \\
BW-f6	&	-0.21	& 	 0.06	&	 0.16	&	 0.08	&	-0.08	&	 0.26 \\
BW-f7	&	 0.11	& 	 0.10	&	-0.10	&	 0.48	&	 0.12	&	 0.02 \\
BW-f8	&	-1.27	& 	 0.66	&	  {NaN}	&	 0.92	&	 0.66	&	 0.48 \\
BWc-01	&	 0.09	& 	 0.20	&	-0.12	&	-0.04	&	-0.08	&	 0.14 \\
BWc-02	&	 0.18	& 	 0.10	&	-0.14	&	-0.24	&	-0.40	&	-0.16 \\
BWc-03	&	 0.28	& 	-0.36	&	-0.34	&	-0.54	&	-0.66	&	-0.22 \\
BWc-04	&	 0.05	& 	-0.08	&	-0.34	&	-0.14	&	-0.30	&	 0.10 \\
BWc-05	&	 0.42	& 	-0.74	&	-0.50	&	{NaN}	&	-0.56	&	-0.64 \\
BWc-06	&	-0.25	& 	 0.20	&	 0.26	&	 0.18	&	 0.26	&	 0.30 \\
BWc-07	&	-0.25	& 	 0.40	&	 0.32	&	 0.32	&	 0.20	&	 0.44 \\
BWc-08	&	 0.37	& 	-0.16	&	-0.22	&	-0.24	&	 0.08	&	-0.16 \\
BWc-09	&	 0.15	& 	 0.08	&	-0.12	&	 0.02	&	-0.24	&	 0.04 \\
BWc-10	&	 0.07	& 	 0.22	&	-0.20	&	 0.32	&	-0.10	&	 0.06 \\
BWc-11	&	 0.17	& 	 0.24	&	-0.06	&	-0.76	&	 0.36	&	 0.10 \\
BWc-12	&	 0.23	& 	 0.14	&	 0.02	&	 0.02	&	-0.02	&	-0.14 \\
BWc-13	&	 0.36	& 	 0.24	&	-0.26	&	-0.16	&	-0.28	&	-0.28 \\
  \hline
\end{tabular}

%% file: Table.type=lines_data.tex
\begin{tabular}{lS[table-format=4.3]S[table-format=1.8]S[table-format=5.3]}
  \hline
  \hline
  \noalign{\smallskip}
  Species & $\lambda$        & {$\chi_{ex}$}           & {$\log gf_{\mathrm{adopted}}$} \\
          & \SI{}{\angstrom} & {\SI{}{\electronvolt}} &                         \\
  \hline
  \noalign{\smallskip}
  \ion{Ba}{II}  & 6496.897  & 0.604321  & -0.320 \\
  \ion{La}{II}  & 6262.287  & 0.403019  & -1.600$^{1}$ \\
  \ion{La}{II}  & 6320.376  & 0.172903  & -1.562 \\
  \ion{La}{II}  & 6390.477  & 0.321339  & -1.410 \\
  \ion{Nd}{II}  & 6740.078  & 0.064     & -1.526 \\
  \ion{Ce}{II}  & 6043.373  & 1.206     & -0.500 \\
  \ion{Eu}{II}  & 6645.064  & 1.379816  &  0.120 \\
  \hline                  
\end{tabular}

%% file: Table.type=abundances_mu_leo.tex
\begin{tabular}{lS[table-format=4.3]S[table-format=1.2]S[table-format=1.2]S[table-format=1.2]S[table-format=1.2]}
  \hline
  \hline
  \noalign{\smallskip}
  Species & {$\lambda$}        & {ESPaDOnS}    & {UVES,$\infty$}    & {UVES, low}   & {UVES, high}\\
          & {\SI{}{\angstrom}} & {\SI{}{\dex}} & {\SI{}{\dex}}      & {\SI{}{\dex}} & {\SI{}{\dex}}\\
  \hline
  \noalign{\smallskip}
  \ion{Ba}{II}  & 6496  & -0.02 &  0.06 & 0.15 $\pm$ 0.09 & 0.10 $\pm$ 0.08\\
  \ion{La}{II}  & 6390 & -0.36 & -0.36 & -0.33 $\pm$ 0.06 & -0.37 $\pm$ 0.03\\
  \ion{Nd}{II}  & 6740  & -0.70 & -0.66 & -0.56 $\pm$ 0.08 & -0.65 $\pm$ 0.04\\
  \ion{Ce}{II}  & 6043  & -0.14 &  0.02 &  0.10 $\pm$ 0.07 &  0.02 $\pm$ 0.04\\
  \ion{Eu}{II}  & 6645  & -0.18 & -0.14 & -0.10 $\pm$ 0.05 & -0.14 $\pm$ 0.03\\
  \hline
\end{tabular}

%% file: Table.type=errors.tex
\begin{tabular}{lcS[table-format=1.2]S[table-format=1.2]S[table-format=1.2]S[table-format=1.2]S[table-format=1.2]S[table-format=1.2]S[table-format=1.2]S[table-format=1.2]}
  \hline
  \hline
  \noalign{\smallskip}
  \multicolumn{2}{c}{Line} & {$\Delta_{+}(T)$} & {$\Delta_{-}(T)$} & {$\Delta_{+}(\log g)$} & {$\Delta_{-}(\log g)$} & {$\Delta_{+}(\abratio{M}{H})$} & {$\Delta_{-}(\abratio{M}{H})$} & {$\Delta_{+}(v_{\text{t}})$} & {$\Delta_{-}(v_{\text{t}})$}\\
       & {\SI{}{\angstrom}} & \multicolumn{8}{c}{\SI{}{\dex}}\\
  \hline
  \noalign{\smallskip}
  Ba & 6497 &	 0.10 &	-0.10 &	 0.08 &	-0.10 &	-0.02 &	 0.00 &	-0.16 &	 0.16\\
  La & 6320 &	 0.04 &	-0.04 &	 0.10 &	-0.10 &	 0.00 &	 0.02 &	 0.00 &	 0.00\\
  Nd & 6740 &	 0.02 &	-0.02 &	 0.08 &	-0.08 &	 0.00 &	 0.00 &	 0.00 &	 0.00\\
  Ce & 6043 &	 0.02 &	-0.02 &	 0.10 &	-0.08 &	 0.02 &	 0.02 &	 0.02 &	 0.00\\
  Eu & 6645 &	 0.00 &	 0.02 &	 0.10 &	-0.08 &	 0.00 &	 0.02 &	 0.00 &	 0.00\\
  \hline 
\end{tabular}

%% file: Table.type=s_elements.tex
\begin{tabular}{lS[table-format=5.2]S[table-format=5.2]S[table-format=5.2]S[table-format=5.2]}
  \hline
  \hline
  \noalign{\smallskip}
           & \multicolumn{2}{c}{$\SI{-0.6}{} < \abratio{Fe}{H} < \SI{0}{}$}        & \multicolumn{2}{c}{$\SI{0}{} < \abratio{Fe}{H}$} \\
  Species  & {$<\abratio{X}{Fe}>$} & {std. dev.} & {$<\abratio{X}{Fe}>$} & {std. dev.}    \\
  \hline
  \noalign{\smallskip}
  \ion{Ba}{II}  & 0.07  & 0.17  & -0.06 & 0.20 \\
  \ion{La}{II}  & 0.16  & 0.19  & -0.06 & 0.20 \\
  \ion{Nd}{II}  & 0.01  & 0.26  & -0.24 & 0.23 \\
  \ion{Ce}{II}  & 0.09  & 0.26  & -0.27 & 0.42 \\
  \hline                  
\end{tabular}